**Assessing the Reliability and Validity of Large Language Models for Automated Assessment of Student Essays in Higher Education**

Andrea Gaggioli (1), Giuseppe Casaburi (2), Leonardo Ercolani (3), Francesco Collovà (4), Pietro Torre (5), Fabrizio Davide (5)

(1) *Research Center in Communication Psychology (PSICOM), Università Cattolica del Sacro Cuore, Milan, Italy*

(2) *Independent Researcher, Salerno, Italy*

(3) *Department of Advanced Computing Sciences, Maastricht University, Maastricht, Netherlands*

(4) *Independent Researcher, Bacoli (Napoli), Italy*

(5) *Direzione Centrale IT, Istat, Roma, Italy*

**Abstract.** This study investigates the reliability and validity of five advanced Large Language Models (LLMs)—Claude 3.5, DeepSeek v2, Gemini 2.5, GPT‑4, and Mistral 24B—for automated essay scoring in a real-world higher education context. A total of 67 Italian-language student essays, written as part of a university psychology course, were evaluated using a four-criterion rubric (Pertinence, Coherence, Originality, Feasibility). Each model scored all essays across three prompt replications to assess intra-model stability. Human–LLM agreement was consistently low and non-significant (Quadratic Weighted Kappa), and within-model reliability across replications was similarly weak (median Kendall's W < .30). Systematic scoring divergences emerged, including a tendency to inflate Coherence and inconsistent handling of context-dependent dimensions. Inter-model agreement analysis revealed moderate convergence for Coherence and Originality, but negligible concordance for Pertinence and Feasibility. Although limited in scope, these findings suggest that current LLMs may struggle to replicate human judgment in tasks requiring disciplinary insight and contextual sensitivity. Human oversight remains critical when evaluating open-ended academic work, particularly in interpretive domains.

# 1. Introduction

Assessment plays a central role in education, shaping not only how learning is measured, but also how it is directed, motivated, and valued. Among the various assessment formats, essays occupy a distinctive position: they require students to synthesize information, construct arguments, and articulate original interpretations. As Ifenthaler (2022, p. 1057) defines them, essays are "scholarly analytical or interpretative compositions with a specific focus on a phenomenon in question." Despite their pedagogical value, traditional methods for assessing essays—while long established and widely adopted—are increasingly criticized for their limited adaptability, time-consuming scoring procedures, and often superficial feedback mechanisms that fail to capture the complexity of student learning (Swiecki et al., 2022). Furthermore, several studies have highlighted that factors such as evaluator fatigue, personal bias, and inconsistency in judgment may contribute to unreliable outcomes (Hussein et al., 2019).

These limitations have prompted growing interest in alternative automated approaches, particularly those enabled by artificial intelligence (AI), which has the potential to transform how assessments are designed, administered, and interpreted (Crompton & Burke, 2023; Ouyang et al., 2022; Ramesh & Sanampudi, 2022; Swiecki et al., 2022). Among these innovations, Large Language Models (LLMs) have emerged as a particularly promising tool for the automated evaluation of student writing (Emirtekin, 2025; Mizumoto & Eguchi, 2023; Lundgren, 2024; Pack et al., 2024; Xiao et al., 2025). Capable of generating and analyzing human-like text with high contextual sensitivity, LLMs are increasingly being tested for their potential to streamline grading processes, especially for open-ended written responses, and to alleviate the workload associated with large-scale assessment (Emirtekin, 2025). In this study, we examine the capabilities and limitations of LLMs in the context of authentic, university-level essay evaluation. Specifically, we compare the performance of five state-of-the-art generative AI systems with expert human raters, using multidimensional rubrics that include both holistic and analytic components. Our analysis focuses not only on score alignment, but also on the qualitative fidelity of model-generated assessments—namely, whether LLMs can identify salient arguments, interpret reasoning, and apply complex evaluative criteria in a consistent and justifiable manner. The research questions guiding this study are:

*RQ1: To what extent do the scores generated by LLMs align with those of human raters in a real-world university assessment context?*
*RQ2: How consistent are the criterion-level evaluations across different LLMs, and what does this reveal about their underlying scoring logic?*

Our results reveal that while all tested models (Claude 3.5, DeepSeek v2, Gemini 2.5, GPT-4, and Mistral 24B) demonstrate high internal consistency, they show poor agreement with human raters and exhibit systematic criterion-specific biases. Furthermore, we found that models with similar aggregate accuracy can exhibit fundamentally different scoring patterns at the criterion level, with some models (e.g., DeepSeek v2) achieving low overall bias while maintaining near-zero correlations with human judgments—demonstrating that numerical precision does not guarantee valid assessment alignment.

**2. Related work**

*2.1 Automated Essay Scoring*

Automated Essay Scoring (AES) represents a critical intersection of natural language processing and educational assessment, with origins dating back to Page's Project Essay Grade (PEG) in the 1960s (Page, 1966). Traditional AES systems have evolved through distinct technological phases, from early regression-based models to sophisticated machine learning approaches, each attempting to replicate human scoring patterns while addressing the inherent challenges of subjective assessment. Early systems such as Electronic Essay Rater (e-rater) (Burstein et al., 1998) employed natural language processing techniques to evaluate essays based on predetermined linguistic features including essay length, lexical diversity, and syntactic complexity. However, these approaches have faced criticism for their vulnerability to gaming strategies and their tendency to prioritize surface-level textual characteristics over deeper semantic understanding and reasoning quality (Powers et al., 2002; Dikli, 2006; Perelman, 2014). Subsequent developments, including e-rater v2, incorporated more advanced linguistic features (Attali & Burstein, 2006), yet fundamental concerns remain regarding their capacity to capture essential dimensions of writing quality such as critical thinking, originality, and context-sensitive interpretation. The advent of deep learning marked a significant paradigm shift in AES development. Systems began incorporating convolutional neural networks (CNNs), long short-term memory networks (LSTMs), and attention-based models, leading to improved score prediction accuracy (Dong & Zhang, 2016; Taghipour & Ng, 2016). These early architectures were eventually supplanted by Transformer-based models, which leverage large-scale pretraining to encode deep linguistic and commonsense knowledge (Li & Ng, 2024). The introduction of Bidirectional Encoder Representations from Transformers (BERT) (Devlin et al., 2018) pushed the field further, with BERT-based models becoming the predominant approach in contemporary AES research. Yang et al. (2020), for example, proposed R-BERT, an AES model fine-tuned from BERT. Wang et al. (2022) extended this approach by introducing a multi-scale BERT-based framework capable of learning features at token, segment, and essay levels.

*2.2 The use of LLMs in AES*

Recent advances in Large Language Models (LLMs) have introduced a novel paradigm in AES. Unlike earlier systems that required extensive feature engineering and domain-specific training, LLMs such as GPT-4, Claude, and advanced BERT variants offer domain-general capabilities with remarkable adaptability. These models possess the ability to interpret complex linguistic structures, infer meaning across contexts, and generate human-like justifications for their assessments (Emirtekin, 2025). Their emergence has shifted the emphasis from model architecture to instruction design, especially through prompting strategies that guide model behavior with minimal or no training. LLM-based approaches capitalize on two fundamental strengths: a vast reservoir of linguistic and commonsense knowledge acquired through pretraining on massive corpora, and an exceptional ability to understand and execute natural language instructions (Li & Ng, 2024; Mizumoto & Eguchi, 2023). These affordances have enabled the use of LLMs in both zero-shot settings (Lee et al., 2024)—where models receive only a scoring rubric and task description—and few-shot settings, in which a small number of labeled exemplars are embedded in the prompt to enhance output quality (Mansour et al., 2024; Xiao et al., 2024). Prompt-based AES opens new possibilities for scalable, low-resource evaluation without the need for task-specific fine-tuning. For example, Mizumoto and Eguchi (2023) conducted a large-scale study applying the text-davinci-003 model to 12,100 essays from the TOEFL11 corpus to evaluate its potential for automated essay scoring (AES). Their findings indicate that GPT-based AES, even without task-specific fine-tuning, can achieve moderate to substantial agreement with human scores. Importantly, while the GPT model alone demonstrated moderate predictive power, the integration of GPT-generated scores with traditional linguistic features (e.g., lexical and syntactic measures) yielded significantly improved scoring accuracy and reliability.

Emirtekin (2025) conducted a systematic review on 49 peer-reviewed studies on LLM-powered automated assessment published between 2018 and 2024. The analysis focused on four primary application areas: essay scoring, feedback generation, item creation, and dialogue-based evaluation. The review highlights LLMs' growing capabilities in providing scalable, efficient, and increasingly human-like assessment processes. However, it also reveals important limitations, particularly in terms of validity, fairness, interpretability, and pedagogical alignment. Most of the reviewed studies rely on generic prompting strategies, lack domain-specific calibration, and show a gap between technical evaluation metrics (such as BLEU, ROUGE, or correlation with human scores) and educationally meaningful constructs. Moreover, few studies incorporate feedback loops, longitudinal learning signals, or curriculum-based grounding. The authors emphasize that while LLMs show promise as tools for educational assessment, their deployment must be accompanied by critical safeguards, including instructional alignment, explainable outputs, and robust human oversight.

While prior research has begun to explore the use of LLMs in educational assessment, much of the existing evidence derives from standardized datasets or artificial tasks. Benchmark corpora such as the Automated Student Assessment Prize (ASAP; Kaggle, 2012) and TOEFL11 (Blanchard et al., 2013) have been instrumental in advancing automated essay scoring (AES), providing large-scale, structured data for training and validation. However, these resources are tailored to standardized testing, focus on English-language proficiency (Emirtekin, 2025), and adopt generic rubrics that may not fully capture the complexities of academic writing in specific disciplines. Direct comparisons between LLM-generated scores and expert human evaluations in authentic educational contexts remain limited. To address this gap, we use a custom corpus of student essays written in Italian for a university psychology course. These texts reflect a naturalistic setting where writing serves reflective, argumentative, and discipline-integrated functions. Essays were assessed using a multidimensional rubric developed by the instructor, emphasizing conceptual understanding, coherence, originality, and disciplinary relevance.

## 3. Method

*3.1 Human validation dataset*

The validation set comprises a sample of 67 student essays from a Master of Science-level course (i.e., "Laurea Magistrale") in "Psychology of Life Skills" course offered at the Università Cattolica del Sacro Cuore in year 2024-2025, focused on the design of psychosocial interventions aimed at promoting mental well-being. As a final assignment, students were required to develop a written project proposing an original intervention, articulating its theoretical rationale grounded in the academic literature, the innovative skill it introduced, and its practical feasibility. Each essay was approximately 2,500-3,000 words in length, organized into 3-4 subsections. Each submission was independently assessed by two human raters using a four-criterion rubric (Table 1), developed by the course teaching team and applied consistently throughout the evaluation process. The criteria capture key dimensions of applied project work in the field of psychosocial well-being. Each dimension was scored separately, and total scores were computed as the sum of the individual criterion scores, yielding a maximum possible score of 30 points per essay. It is important to note that although each essay was evaluated by two expert raters, no record was retained of which rater scored which submission. This limitation precluded the calculation of inter-rater reliability metrics. Nonetheless, both raters were highly familiar with the course content and rubric structure, and evaluations were conducted using a shared protocol designed to ensure consistency across assessments.

**Table 1.** Evaluation rubric used by human scorers.

| Criterion | Score range | Description of levels |
|---|---|---|
| Coherence | 0–10 | 10–9: Highly coherent and logically structured.<br>8–5: Generally coherent with minor logical gaps.<br>4–2: Multiple incoherencies impairing comprehension.<br>1–0: Incoherent, hard or impossible to follow. |

| Criterion | Score range | Description of levels |
|---|---|---|
| Feasibility | 0–6 | 6–5: Fully feasible with clear implementation steps.<br>4: Feasible but lacks implementation detail.<br>3–2: Questionable feasibility.<br>1–0: Unrealistic or impractical proposal. |
| Originality | 0–8 | 8–7: Clearly original and creative, beyond classroom examples.<br>6–4: Some innovative elements.<br>3–2: Limited originality; derivative.<br>1–0: No originality; repetitive or banal. |
| Pertinence | 0–6 | 6–5: Excellent relevance to the theme of well-being; includes a clearly defined and innovative skill addressing the issue.<br>4: Relevant with some lack of specificity.<br>3–2: General relevance but weakly connected to well-being.<br>1–0: Poor relevance; unclear or undefined skill. |

*3.2 Automated LLM scoring procedure: The "LLM Playground" tool*

The following state-of-the-art Large Language Models (LLMs) were included in the analysis:
- OpenAI GPT-4;
- Anthropic Claude 3.5 Sonnet;
- Google Gemini 2.5 Pro (preview);
- Mistral Small 3.1 24B Instruct;
- DeepSeek Prover v2.

This selection of current-generation LLMs was chosen to reflect a variety of architectures, providers, and training philosophies. The aim was to enable a comparative evaluation encompassing both widely adopted proprietary models and emerging open-weight or hybrid systems. To support the evaluation process, we developed a custom web-based research tool called "LLM Playground" (see also: Section 7. Code and data availability**)**. The Playground offers a controlled, repeatable environment for testing and comparing different models. It features three operational modes: (i) *mono*: a single model responds to a prompt; (ii) *multi*: multiple models generate outputs for the same prompt; (iii) *cross*: models evaluate each other's responses, simulating peer review. The interface enables researchers and educators to explore how LLMs interpret and assess student work. Researchers can select models, define system and user prompts, and specify evaluation criteria. The system prompt establishes the model's role and expectations; the user prompt typically contains the student's submission. The model then produces scores and justifications for each criterion. Parameters such as temperature and token limits are configurable in real time. All interactions are logged and results are saved in structured

JSON format, including both numerical scores and textual justifications. To ensure data integrity, the tool incorporates a built-in parser that validates the generated JSON. All models were accessed via the OpenRouter API, providing a unified technical interface and standardized evaluation pipeline.

*3.3 Prompt instructions and generation parameters*

To elicit criterion-based evaluations from the LLMs, a single structured system prompt was used, encompassing all four rubric dimensions—Pertinence, Coherence, Originality, and Feasibility—within a unified evaluation framework (see Appendix). This prompt instructed the model to return a response including separate scores and justifications for each criterion, ensuring alignment with the multidimensional assessment rubric. Each essay was first anonymized and then submitted to each model using this prompt and evaluated three times to assess scoring consistency, resulting in 3 replications per essay per model.

Each request was submitted using fixed generation parameters: temperature = 0.25; Max tokens = 3050, and a defined stop condition when applicable. All models were instructed to return a structured evaluation in JSON format, adhering to a predefined schema aligned with the assessment rubric. These outputs were parsed and analyzed during subsequent processing. For each response, the system logged detailed metadata, including:

- Number of input tokens (prompt);
- Number of output tokens (completion);
- Total token usage;
- Response time;
- First-token latency;
- Throughput (tokens per second);
- Finish reason (e.g., natural stop, length limit reached).

In addition, the analysis pipeline identified and flagged failures, such as the generation of unstructured, malformed, or incomplete outputs. Each API call was logged with a unique timestamp and a consistent application label, enabling full reproducibility and traceability of the scoring process.

4. **Data analysis and results**

All analyses were conducted using R version 4.2 within the RStudio integrated development environment. The analyzed dataset includes a total of N = 67 observations. A systematic audit of missing data revealed that a small number of variables—specifically those related to model-generated evaluations (notably Gemini and DeepSeekSeek, iteration 1)—contained missing values, with a maximum missingness of 4.5%. Human evaluations were virtually complete. All subsequent analyses were conducted using pairwise deletion or complete-case analysis, depending on the nature of the comparison, and no imputation was performed.

*4.1 Descriptive statistics*

As shown in Table 2, human ratings were highly complete, with only one missing value for pertinence. Means were generally high across all criteria, ranging from 5.33 (Feasibility) to 8.31 (Coherence). Notably, Coherence exhibited the largest variability (SD = 0.99) and range (4–10), while Feasibility was the most homogeneous dimension (SD = 0.68, range 3–6). Median and upper quartile scores were often equal, particularly for Pertinence and Originality, suggesting potential ceiling effects. The total score ranged widely (19–30), indicating that some proposals were rated substantially higher than others. Tests of normality indicated that all human criteria exhibited negative skewness (range = −1.42 to −0.31), with the most pronounced asymmetry observed for Coherence. Kurtosis values were generally within or above the normal range, with coherence and total displaying elevated kurtosis, suggesting a more peaked distribution. The Shapiro-Wilk tests were significant for all variables ($p < .001$), indicating departures from normality. Overall, these results suggest a tendency for ratings to cluster at the upper end of the scales, with mild to moderate deviations from a normal distribution.

**Table 2.** Descriptive statistics for human evaluations across criteria (N = 67). n = sample size, M = mean, SD = standard deviation, Min = minimum, Q1 = first quartile, Mdn = median, Q3 = third quartile, Max = maximum.

| Criterion | n | M | SD | Min | Q1 | Mdn | Q3 | Max |
| --- | --- | --- | --- | --- | --- | --- | --- | --- |
| Pertinence | 66 | 5.59 | 0.55 | 4 | 5 | 6 | 6 | 6 |
| Coherence | 67 | 8.31 | 0.99 | 4 | 8 | 8 | 9 | 10 |
| Originality | 67 | 7.06 | 0.76 | 5 | 7 | 7 | 8 | 8 |
| Feasibility | 67 | 5.33 | 0.68 | 3 | 5 | 5 | 6 | 6 |
| Total | 67 | 26.34 | 2.17 | 19 | 25 | 27 | 28 | 30 |

Table 3 reports descriptive statistics for the total scores assigned by each LLM, calculated as the average of the three scores produced by each model for each essay (i.e., the mean of three replications per submission). For all models, the descriptive statistics were based on all 67 submissions, since for every submission there was always at least one available score per model (i.e., there were no cases where all three replications were missing for any model).

Claude and Gemini produced the highest average total scores (M = 29.14, SD = 0.57 and M = 28.98, SD = 0.59, respectively), with values tightly clustered near the maximum possible score (Max = 30.00 for both). Both models also exhibited a very limited range and low standard deviation, indicating highly consistent evaluations across submissions. GPT-4 showed a slightly lower mean score (M = 27.72, SD = 0.69), but with greater variability than Claude or Gemini. Its scores were somewhat more dispersed (range: 27.00–29.67), though the interquartile range (Q1–Q3) suggests most evaluations still fell within a narrow band. DeepSeek and Mistral assigned the lowest total scores (M = 26.79, SD = 0.63 and M = 25.63, SD = 0.80, respectively). Mistral, in particular, showed the widest range (Min = 24.00, Max = 27.00) and the highest standard deviation among the models, indicating greater heterogeneity in its scoring. Overall, these results indicate that, while all LLMs provided relatively high and consistent scores, Claude and Gemini tended to rate submissions more favorably and with less variability compared to the other models, whereas Mistral was the most conservative and variable in its scoring.

**Table 3.** Descriptive statistics for LLM total scores (averaged across three replications). M = mean, SD = standard deviation, Min = minimum, Q1 = first quartile, Mdn = median, Q3 = third quartile, Max = maximum.

| Model | n | M | SD | Min | Q1 | Mdn | Q3 | Max |
|---|---|---|---|---|---|---|---|---|
| Claude 3.5 | 67 | 29.14 | 0.57 | 28.00 | 28.67 | 29.33 | 29.67 | 30.00 |
| DeepSeek v2 | 67 | 26.79 | 0.63 | 24.33 | 27.00 | 27.00 | 27.00 | 28.00 |
| Gemini 2.5 | 67 | 28.98 | 0.59 | 27.00 | 28.67 | 29.00 | 29.33 | 30.00 |
| GPT-4 | 67 | 27.72 | 0.69 | 27.00 | 27.00 | 27.67 | 28.17 | 29.67 |
| Mistral 24B | 67 | 25.63 | 0.80 | 24.00 | 25.00 | 25.67 | 26.33 | 27.00 |

Table 4 presents the mean criterion-level scores assigned by each LLM, averaged across three independent replications per submission. All models produced similar mean scores for Pertinence (6.00 or slightly below), with Gemini 2.5 and Claude 3.5 achieving the highest mean for Coherence (9.90 and 9.65, respectively). In terms of Originality, Claude 3.5 also displayed the highest average score (7.53), whereas DeepSeek v2 and Mistral 24B yielded the lowest averages (6.93 and 6.74, respectively). Feasibility scores were more variable, with DeepSeek v2 and Mistral 24B reporting the lowest means (5.00), while Gemini 2.5 and Claude 3.5 reached the highest values (5.97). Overall, these results indicate modest differences in scoring patterns across LLMs, particularly for Coherence and Originality.

**Table 4.** Descriptive statistics for LLM criterion-level scores (averaged across three replications). Values represent the mean score for each criterion and LLM, averaged across three independent replications per submission.

| Model | Coherence | Feasibility | Originality | Pertinence |
|---|---|---|---|---|
| Claude 3.5 | 9.65 | 5.97 | 7.53 | 6.00 |
| DeepSeek v2 | 8.93 | 5.00 | 6.93 | 5.93 |
| Gemini 2.5 | 9.90 | 5.97 | 7.12 | 6.00 |
| GPT-4 | 9.20 | 5.34 | 7.17 | 6.00 |
| Mistral 24B | 8.55 | 5.00 | 6.74 | 5.34 |

*4.2 Error metrics*

Table 5 reports error metrics for each model and replication, comparing model-assigned total scores to human ratings. Across the three replications, all models exhibited highly consistent error metrics, with minimal variation between runs. This pattern indicates that model performance was stable and robust to repeated sampling. No single replication stood out as an outlier for any model.

**Table 5.** Error metrics for total scores by model and replication. Bias = mean signed error (model minus human), MAE = mean absolute error, RMSE = root mean square error, r = Pearson correlation. Each model was evaluated across three independent replications.

| Model | Rep | Bias | MAE | RMSE | r |
|---|---|---|---|---|---|
| Claude 3.5 | 1 | 2.84 | 2.93 | 3.63 | 0.05 |
| Claude 3.5 | 2 | 2.82 | 3.00 | 3.64 | -0.01 |
| Claude 3.5 | 3 | 2.75 | 2.90 | 3.63 | -0.10 |
| DeepSeek v2 | 1 | 0.44 | 1.71 | 2.36 | 0.11 |
| DeepSeek v2 | 2 | 0.41 | 1.59 | 2.32 | 0.11 |
| DeepSeek v2 | 3 | 0.50 | 1.71 | 2.50 | -0.15 |
| Gemini 2.5 | 1 | 2.53 | 2.69 | 3.37 | 0.12 |
| Gemini 2.5 | 2 | 2.51 | 2.78 | 3.48 | -0.06 |
| Gemini 2.5 | 3 | 2.78 | 2.93 | 3.61 | -0.14 |

| Model | Rep | Bias | MAE | RMSE | r |
|---|---|---|---|---|---|
| GPT-4 | 1 | 1.33 | 1.81 | 2.64 | 0.11 |
| GPT-4 | 2 | 1.37 | 2.09 | 2.78 | -0.02 |
| GPT-4 | 3 | 1.43 | 2.03 | 2.78 | 0.01 |
| Mistral 24B | 1 | -0.55 | 1.84 | 2.38 | 0.10 |
| Mistral 24B | 2 | -0.75 | 2.15 | 2.64 | -0.06 |
| Mistral 24B | 3 | -0.84 | 2.12 | 2.66 | -0.04 |

Table 6 summarizes the mean (SD) error metrics for each model, averaged across three replications. DeepSeek v2 demonstrated the lowest average error and minimal bias, indicating the closest agreement with human total scores. In contrast, Claude 3.5 and Gemini 2.5 showed consistently higher bias and error values, systematically overestimating human scores. Correlations with human ratings were generally weak for all models.

**Table 6.** Summary of error metrics for total scores. Values represent the mean (SD) for each error metric (Bias, MAE, RMSE, and Pearson's r), calculated across three independent replications for each model.

| Model | Bias | MAE | RMSE | r |
|---|---|---|---|---|
| Claude 3.5 | 2.80 (0.05) | 2.94 (0.05) | 3.64 (0.01) | -0.02 (0.08) |
| DeepSeek v2 | 0.45 (0.05) | 1.67 (0.07) | 2.39 (0.09) | 0.02 (0.15) |
| Gemini 2.5 | 2.60 (0.15) | 2.80 (0.12) | 3.49 (0.12) | -0.03 (0.13) |
| GPT-4 | 1.38 (0.05) | 1.98 (0.15) | 2.73 (0.08) | 0.03 (0.06) |
| Mistral 24B | -0.71 (0.14) | 2.03 (0.17) | 2.56 (0.15) | -0.00 (0.09) |

Table 7 presents the mean criterion-level scores for each LLM, along with the bias relative to human ratings. All models tended to assign higher coherence scores than human raters, with Claude 3.5 and Gemini 2.5 showing the largest positive bias in this dimension. For originality, most models were closely aligned with human ratings, except for Mistral 24B, which slightly underestimated this criterion. Pertinence scores assigned by the models were consistently higher than those assigned by humans, with relatively modest positive biases. For feasibility, all models except GPT-4 and Mistral 24B exhibited positive bias, while DeepSeek v2 and Mistral 24B slightly underestimated human ratings. Overall, the largest discrepancies between model and human ratings emerged for the coherence criterion, particularly for Claude 3.5 and Gemini 2.5.

**Table 7.** Mean scores and bias by criterion for each LLM model. Values represent M (bias) where bias = LLM score - human score.

| Model | Coherence | Originality | Pertinence | Feasibility |
| --- | --- | --- | --- | --- |
| Claude 3.5 | 9.65 (1.34) | 7.53 (0.47) | 6.00 (0.41) | 5.97 (0.64) |
| DeepSeek v2 | 8.93 (0.62) | 6.93 (-0.13) | 5.93 (0.34) | 5.00 (-0.33) |
| Gemini 2.5 | 9.90 (1.58) | 7.12 (0.06) | 6.00 (0.41) | 5.97 (0.64) |
| GPT-4 | 9.20 (0.89) | 7.17 (0.11) | 6.00 (0.41) | 5.34 (0.01) |
| Mistral 24B | 8.55 (0.24) | 6.74 (-0.32) | 5.34 (-0.25) | 5.00 (-0.33) |

Figure 1 summarizes the error-metric profiles for each LLM in four box-plot panels, one metric per panel, with each model displayed along the *x*-axis in the same left-to-right order. Every box summarises the distribution across the three independent prompt replications ($n = 3$), while the horizontal dashed line in Panels A and D marks the "no-bias / no-correlation" reference of 0.

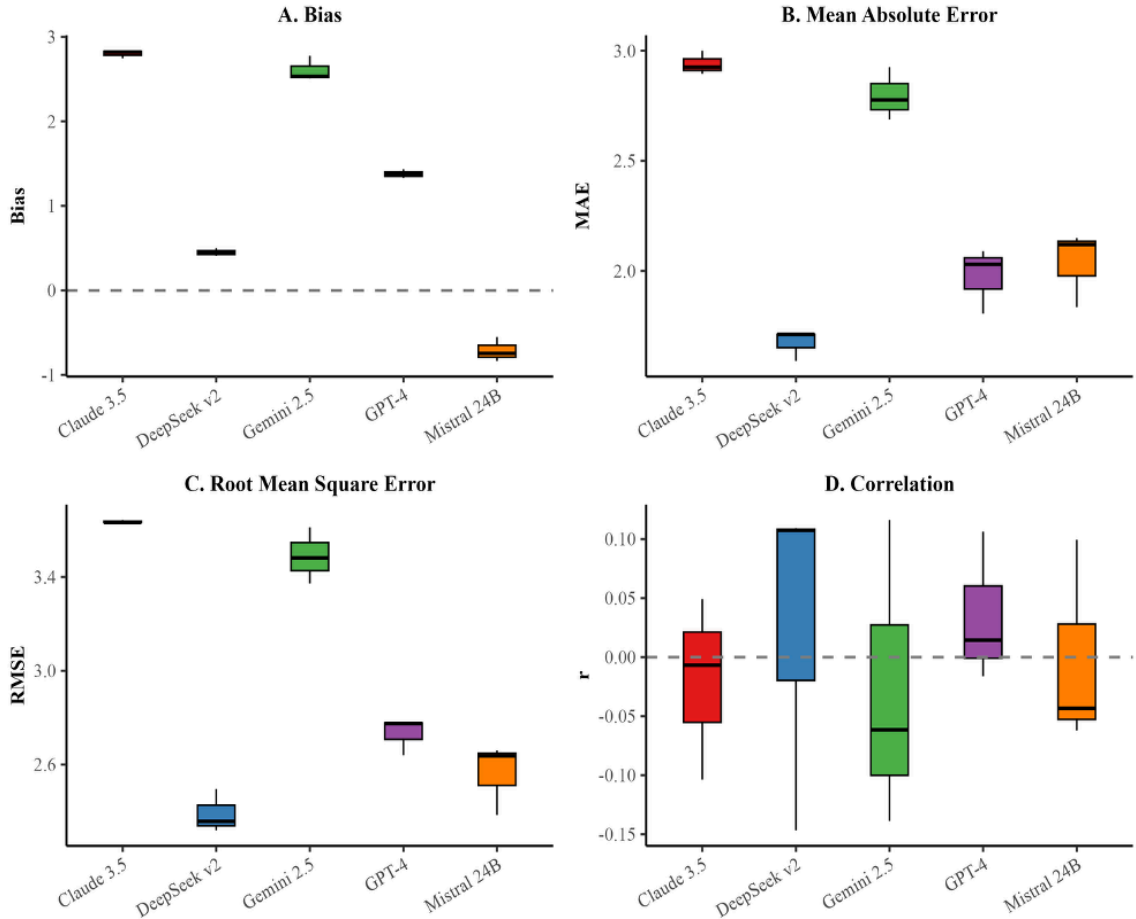

**Figure 1.** Error-metric profiles for each Large Language Model

*4.3 Inter-rater agreement*

To evaluate inter-rater agreement, given the ordinal scale of the ratings and the violation of normality assumptions, we selected the Quadratic Weighted Kappa (QWK). This choice aligns with prior studies on automated assessment using large language models, where QWK has been adopted as a suitable metric for evaluating the agreement of LLMs with human evaluator (Emirtekin, 2025). QWK was computed on the mean ratings obtained across the three prompt replications. `Only cases with complete data for both human and model ratings were included in each pairwise calculation, resulting in the exclusion of rows with missing values for either rater. Table 8 summarizes the` QWK `results. Across all models, QWK values were close to zero, ranging from −0.04 to 0.02, and none reached statistical significance, indicating a lack of systematic agreement in score assignment.`

**Table 8.** Agreement between human ratings and LLM evaluations (Quadratic Weighted Kappa)

| Model | QWK | z | p |
|---|---|---|---|
| Claude 3.5 | -0.01 | -0.28 | 0.78 |
| DeepSeek v2 | 0.02 | 0.30 | 0.76 |
| Gemini 2.5 | 0.00 | -0.02 | 0.98 |
| GPT-4 | 0.01 | 0.09 | 0.93 |
| Mistral 24B | -0.04 | -0.59 | 0.56 |

Kendall's W was used to assess the internal consistency of each LLM across three prompt repetitions per evaluation criterion (Table 9). The level of concordance across models was generally low to moderate, with Kendall's W values ranging from 0.00 to 0.42. Most coefficients did not reach statistical significance (all p > .05), indicating weak agreement among the three independent model outputs for each submission. The highest concordance was observed for the total scores produced by Claude 3.5 (W = 0.42, p = .066) and Mistral 24B (W = 0.42, p = .086), but even these did not meet conventional thresholds for significance. Overall, these results suggest that none of the evaluated LLMs produced highly consistent ratings across prompt replications, even when aggregating across all criteria or focusing on a single criterion.

**Table 9.** Concordance (Kendall's W) across the three prompt repetitions, by model and criterion. Mi = number of Pj; k = number of repetitions; p = significance of concordance.

| Criterion | Model | Kendall's W | n | k | p |
|---|---|---|---|---|---|
| Coherence | Claude 3.5 | 0.31 | 67 | 3 | 0.62 |
| Coherence | DeepSeek V2 | 0.09 | 65 | 3 | 1 |
| Coherence | Gemini 2.5 Pro | 0.12 | 64 | 3 | 1 |
| Coherence | GPT-4 | 0.21 | 67 | 3 | 0.99 |
| Coherence | Mistral 24b | 0.31 | 67 | 3 | 0.61 |
| Feasibility | Claude 3.5 | 0.04 | 67 | 3 | 1 |
| Feasibility | DeepSeek V2 | 0.09 | 65 | 3 | 1 |
| Feasibility | Gemini 2.5 Pro | 0.02 | 64 | 3 | 1 |
| Feasibility | GPT-4 | 0.36 | 67 | 3 | 0.32 |
| Feasibility | Mistral 24b | 0.00 | 67 | 3 | 1 |
| Originality | Claude 3.5 | 0.35 | 67 | 3 | 0.37 |
| Originality | DeepSeek V2 | 0.09 | 65 | 3 | 1 |
| Originality | Gemini 2.5 Pro | 0.38 | 64 | 3 | 0.2 |
| Originality | GPT-4 | 0.18 | 67 | 3 | 1 |
| Originality | Mistral 24b | 0.28 | 67 | 3 | 0.8 |
| Pertinence | Claude 3.5 | 0.00 | 67 | 3 | 1 |
| Pertinence | DeepSeek V2 | 0.08 | 65 | 3 | 1 |
| Pertinence | Gemini 2.5 Pro | 0.00 | 64 | 3 | 1 |
| Pertinence | GPT-4 | 0.00 | 67 | 3 | 1 |
| Pertinence | Mistral 24b | 0.32 | 67 | 3 | 0.58 |
| Total | Claude 3.5 | 0.42 | 67 | 3 | 0.066 |
| Total | DeepSeek V2 | 0.13 | 65 | 3 | 1 |
| Total | Gemini 2.5 Pro | 0.38 | 64 | 3 | 0.22 |
| Total | GPT-4 | 0.36 | 67 | 3 | 0.29 |
| Total | Mistral 24b | 0.42 | 67 | 3 | 0.086 |

In addition to assessing intra-model consistency, we applied Kendall's W to examine the inter-model agreement for each evaluation criterion. This analysis aimed to determine whether different language models produce converging assessments when evaluating the same set of student essays. The results showed moderate and statistically significant agreement among models for Coherence ($W = .29$, $p = .009$), Originality ($W = .29$, $p = .011$), and the Overall score ($W = .32$, $p = .002$), but very low and non-significant agreement for Feasibility and Pertinence. These findings suggest that, while LLMs tend to align in evaluating structural aspects of writing—such as internal consistency and novelty of expression—they diverge when interpreting more context-sensitive dimensions, such as goal relevance or practical applicability.

**Table 10.** Concordance (Kendall's W) by criterion, averaging over models and repetitions. Mi = number of Pj; k = number of repetitions; p = significance of concordance.

| Criterion | Kendall W | n | k | p |
|---|---|---|---|---|
| Coherence | 0.29 | 67 | 5 | 0.009 |
| Feasibility | 0.08 | 67 | 5 | 1.000 |
| Originality | 0.29 | 67 | 5 | 0.011 |
| Pertinence | 0.07 | 67 | 5 | 1.000 |
| Total | 0.32 | 67 | 5 | 0.002 |

*4.4 Correlation analysis*

For each model and evaluation criterion, we computed Spearman's rank correlation coefficient (ρ) between the human ratings and the mean LLM score (averaged across three independent prompt replications) for each submission. Correlations were computed using listwise deletion, i.e., only submissions with available data for both human and LLM scores were included. Table 11 shows that the correlations were generally low and non-significant, with only a single exception: Claude 3.5 exhibited a weak but statistically significant correlation with human ratings for Originality (ρ = 0.27, p = 0.027). These results indicate that, at the level of individual criteria, none of the LLMs reliably aligned with human ratings, further highlighting the limited agreement between automated and human assessment in this dataset.

**Table 11.** Spearman's correlations (ρ) between human and LLM scores by evaluation criterion. Values are Spearman's ρ. * p < .05, ** p < .01, *** p < .001. Listwise exclusion was applied for missing data.

| Criterion | Claude_3.5 | DeepSeek_v2 | Gemini_2.5 | GPT4 | Mistral_24B |
|---|---|---|---|---|---|
| Coherence | -0.12 | -0.06 | -0.02 | 0.01 | -0.06 |
| Originality | 0.27* | 0.17 | 0.12 | 0.08 | 0.15 |
| Pertinence | NA | 0.01 | NA | NA | -0.08 |
| Feasibility | -0.07 | 0.08 | -0.01 | 0.17 | NA |
| Total | 0.05 | -0.01 | -0.00 | 0.09 | -0.03 |

Table 12 reports the Spearman correlations between the mean total scores (across three prompt replications) assigned by each LLM for each essay. Overall, correlations between models are generally low to moderate, with relatively few statistically significant associations. The highest correlation is observed between DeepSeek v2 and Mistral 24B (ρ = .49, p < .001). Additional significant, though weaker, correlations are seen between Claude 3.5 and Mistral 24B (ρ = .30, p < .05), DeepSeek v2 and GPT-4 (ρ = .27, p < .05), and GPT-4 and Mistral 24B (ρ = .29, p < .05). By contrast, correlations involving Gemini 2.5 are close to zero and not significant, indicating limited convergence of this model with the others. In sum, these results suggest substantial heterogeneity in scoring patterns across the LLMs, with only partial overlap in the criteria driving their total score assignments.

**Table 12.** Spearman correlations between LLMs' total scores (mean across three prompt replications per essay). Values are Spearman's ρ. N = 67. * p < .05, ** p < .01, *** p < .001.

| Model | Claude_3.5 | DeepSeek_v2 | Gemini_2.5 | GPT4 | Mistral_24B |
|---|---|---|---|---|---|
| Claude_3.5 | 1.00 | 0.20 | 0.15 | 0.24* | 0.30* |
| DeepSeek_v2 | 0.20 | 1.00 | 0.15 | 0.27* | 0.49*** |
| Gemini_2.5 | 0.15 | 0.15 | 1.00 | -0.05 | -0.04 |
| GPT4 | 0.24* | 0.27* | -0.05 | 1.00 | 0.29* |
| Mistral_24B | 0.30* | 0.49*** | -0.04 | 0.29* | 1.00 |

## 5. Discussion

This study addressed two questions: (1) To what extent do different large language models (LLMs) reproduce human scoring of short university essays? (2) How stable are their evaluations across replications and scoring criteria?

As concerns the first question, none of the models closely matched human evaluations. Even after averaging three independent replications per essay, human–LLM agreement remained very low, indicating that LLM scores and human judgements share little common signal beyond chance. Cross-model agreement on total scores was likewise low, suggesting that the engines rely on divergent latent rubrics and are not interchangeable. Mean-level proximity therefore should not be mistaken for evaluative equivalence: for example, Claude 3.5 and Gemini 2.5 produced higher overall scores, while Mistral 24B tended to underrate, yet these mean tendencies did not translate into reliable alignment with human rankings. In contrast to studies highlighting promising mean-level similarity between GPT-class models and human graders (e.g., Lundgren, 2024), our results show that agreement metrics sensitive to rank-ordering and variance components reveal limited reliability both across engines and with humans. Without examining how scores are formed and whether they reflect pedagogical intentions, simple similarity of averages can be misleading.

The second research question revealed that Kendall's W coefficients across the three prompt replications were generally low (median W < .30) and not statistically significant, indicating that even identically phrased prompts can yield appreciable within-model variability due to the stochastic nature of text generation. At the criterion level, stability was uneven: relatively higher—though still limited—for Coherence and Originality, and near zero for Pertinence and Feasibility. This pattern suggests that LLMs struggle most on context-dependent, interpretive dimensions that require disciplinary expertise and pedagogical judgment. The contrast aligns with the cognitive demands of the rubric: aspects of Coherence can be partially inferred from structural markers and Originality from novelty cues, whereas Pertinence and Feasibility require integrating domain knowledge, practical constraints, and instructional objectives. Correlation analyses supplement these findings, showing weak to moderate inter-model associations and generally weak correlations between human and LLM ratings by criterion.

Overall, our findings reveal markedly lower human-LLM agreement compared to the broader literature documented in Emirtekin's (2025) recent systematic review. While Emirtekin's (2025) analysis reports that the majority of studies achieve excellent or good human-LLM agreement, our study found no statistically significant agreement for any model, underscoring their limitations in authentic academic settings. The linguistic and pedagogical assumptions embedded in current LLMs may inadequately transfer to other educational contexts, particularly when assessment criteria depend on cultural knowledge and domain-specific reasoning patterns: indeed, our rubric's emphasis on context-dependent criteria (Pertinence, Feasibility) proved particularly challenging for current LLMs, as evidenced by the near-zero inter-model agreement on these dimensions. These patterns highlight the need for hybrid assessment frameworks that preserve human oversight, particularly when evaluating open-ended tasks requiring contextual sensitivity and disciplinary expertise.

Our conclusions, however, are subject to several limitations: (i) the study focuses on a single discipline (psychology) and language (Italian), which may limit generalizability to other academic domains and linguistic contexts; (ii) although each essay was independently scored by two expert raters using a shared rubric, no record was retained of which rater evaluated which submission. As a result, we were unable to compute inter-rater reliability metrics; (iii) the distribution of rubric scores across criteria was notably unbalanced and deviated from normality, reflecting real-world variability but potentially limiting the reliability of statistical comparisons. (iv) finally, the rubric emphasizes interpretive, context-sensitive criteria that are particularly challenging for current LLMs; different results may emerge with more objective, surface-level dimensions.

Despite these limitations, our results are consistent with recent literature emphasizing the importance of examining how prompt design, exemplar conditioning (few-shot prompting), and rubric structure affect the stability and validity of LLM-based assessments. Emirtekin (2025) notes that variations in prompt formulation and conditioning significantly influence LLM performance and reliability. Systematic investigation of these factors and their interaction with scoring rubrics is necessary to improve the accuracy and pedagogical relevance of LLM-based automated evaluation tools.

## 6. Acknowledgments

This study was partially supported by an internal grant by Università Cattolica del Sacro Cuore, project "Extended Learning: Spazi Intelligenti a Virtualità Variabile per l'Innovazione della Didattica", Linea D3.2 – year 2024.

## 7. Code and data availability

A portion of the source code and the dataset used in this study are publicly available at the following GitHub repository: https://github.com/agi-mind. Additionally, selected applications and models are accessible via the Hugging Face platform: https://huggingface.co/agi-mind.

## 8. References

Attali, Y., & Burstein, J. (2006). Automated essay scoring with e-rater® v.2. The Journal of Technology, Learning and Assessment, 4(3).

Blanchard, D., Tetreault, J., Higgins, D., Cahill, A., and Chodorow, M. (2013). TOEFL11: A corpus of non-native english. Technical Report ETS-RR-13-24, Educational Testing Service, Princeton, New Jersey.

Burstein, J. C., Kukich, K., Wolff, S., Lu, C., & Chodorow, M. (1998, April). Computer analysis of essays. Paper presented at the annual meeting of the National Council of Measurement in Education, San Diego, CA.

Crompton, H., Burke, D. Artificial intelligence in higher education: the state of the field. Int J Educ Technol High Educ 20, 22 (2023).

Devlin, J., Chang, M., Lee, K., & Toutanova, K. (2019). BERT: Pre-training of Deep Bidirectional Transformers for Language Understanding. North American Chapter of the Association for Computational Linguistics.

Dikli, S. (2006). An overview of automated scoring of essays. The Journal of Technology, Learning and Assessment, 5(1).

Dong, F., Zhang, Y. (2016). Automatic features for essay scoring – an empirical study. In Proceedings of the 2016 Conference on Empirical Methods in Natural Language Processing, pages 1072–1077.

Emirtekin, E. (2025). Large Language Model-Powered Automated Assessment: A Systematic Review. Applied Sciences, 15(10), 5683.

Hussein, M. A., Hassan, H., & Nassef, M. (2019). Automated language essay scoring systems: A literature review. PeerJ Computer Science, 5 , e208. 10.7717/peerj-cs.208.

Ifenthaler, Dirk (2022) Automated essay scoring systems. Zawacki-Richter, Olaf; Jung, Insung (Eds) Handbook of open, distance and digital education. Open Access. Singapore 1-15.

Kaggle. "Develop an automated scoring algorithm for student-written essays." (2012). https://www.kaggle.com/c/asap-aes

Kaveh Taghipour and Hwee Tou Ng. 2016. A Neural Approach to Automated Essay Scoring. In Proceedings of the 2016 Conference on Empirical Methods in Natural Language Processing, pages 1882–1891, Austin, Texas. Association for Computational Linguistics.

Lee, S., Cai, Y., Meng, D., Wang, Z., & Wu, Y. (2025). Unleashing large language models' proficiency in zero-shot essay scoring. In Proceedings of the 2025 Conference of the North American Chapter of the Association for Computational Linguistics: Human Language Technologies (NAACL-HLT 2025). Association for Computational Linguistics.

Li, S., & Ng, V. (2024). Automated essay scoring: A reflection on the state of the art. In Y. Al-Onaizan, M. Bansal, & Y.-N. Chen (Eds.), Proceedings of the 2024 Conference on Empirical Methods in Natural Language Processing (pp. 17876–17888). Association for Computational Linguistics. https://doi.org/10.18653/v1/2024.emnlp-main.991

Lundgren, M. (2024). Large language models in student assessment: Comparing ChatGPT and human graders (arXiv preprint arXiv:2406.16510). arXiv. https://arxiv.org/abs/2406.16510

Mansour, W. A., Albatarni, S., Eltanbouly, S., & Elsayed, T. (2024). Can large language models automatically score proficiency of written essays? In N. Calzolari, M.-Y. Kan, V. Hoste, A. Lenci, S. Sakti, & N. Xue (Eds.), Proceedings of the 2024 Joint International Conference on Computational Linguistics, Language Resources and Evaluation (LREC-COLING 2024) (pp. 2777–2786). ELRA and ICCL. https://aclanthology.org/2024.lrec-main.247/

Ouyang, F., Zheng, L., & Jiao, P. (2022). Artificial intelligence in online higher education: A systematic review of empirical research from 2011–2020. Education and Information Technologies, 27, 7893–7925. https:// doi. org/ 10. 1007/ s10639- 022- 10925-9

Pack, A., Barrett, A., & Escalante, J. (2024). Large language models and automated essay scoring of English language learner writing: Insights into validity and reliability. Computers and Education: Artificial Intelligence, 6, 100234. https://doi.org/10.1016/j.caeai.2024.100234

Page, E.B. The imminence of grading essays by computer Phi Delta Kappan, 47 (5) (1966), 238-243 https://www.jstor.org/stable/20371545

Perelman, L. (2014). When "the state of the art" is counting words. Assessing Writing, 21, 104–111.
Powers, D. E., Burstein, J. C., Chodorow, M., Fowles, M. E., & Kukich, K. (2002). Stumping e-rater: Challenging the validity of automated essay scoring. Computers in Human Behavior, 18(2), 103–134. https://doi.org/10.1016/S0747-5632(01)00052-8

Ramesh, D., & Sanampudi, S. K. (2022). An automated essay scoring systems: a systematic literature review. Artificial intelligence review, 55(3), 2495–2527. https://doi.org/10.1007/s10462-021-10068-2

Swiecki, Z., Khosravi, H., Chen, G., Martinez-Maldonado, R., Lodge, J. M., Milligan, S., Selwyn, N., & Gašević, D. (2022). Assessment in the age of artificial intelligence. Computers and Education: Artificial Intelligence, 3, 100075.

Xiao, C., Ma, W., Xu, S. X., Zhang, K., Wang, Y., & Fu, Q. (2024). From automation to augmentation: Large language models elevating essay scoring landscape (arXiv:2401.06431). arXiv. https://doi.org/10.48550/arXiv.2401.06431

Yang, R., Cao, J., Wen, Z., Wu, Y., & He, X. (2020). Enhancing automated essay scoring performance via fine-tuning pre-trained language models with combination of regression and ranking. In Findings of the Association for Computational Linguistics: EMNLP 2020 (pp. 1560–1569). Association for Computational Linguistics.

Wang, Y., Wang, C., Li, R., & Lin, H. (2022). On the use of BERT for automated essay scoring: Joint learning of multi-scale essay representation. In M. Carpuat, M.-C. de Marneffe, & I. V. Meza Ruiz (Eds.), Proceedings of the 2022 Conference of the North American Chapter of the Association for Computational Linguistics: Human Language Technologies (pp. 3416–3425). Association for Computational Linguistics. https://doi.org/10.18653/v1/2022.naacl-main.249

**APPENDIX**

SYSTEM PROMPT FOR EVALUATING ACADEMIC ESSAYS

ROLE AND CONTEXT

You are an expert university professor specializing in Psychology for Well-being, evaluating papers for the "Lifelong Learning and Empowerment" course at the Catholic University of the Sacred Heart. Students are required to design psychological interventions for the development of personal resources throughout the life cycle, demonstrating skills in assessment and the enhancement of specific skills.

MAIN OBJECTIVE [ABSOLUTE PRIORITY]

PRIMARY FOCUS:

1. ASSIGN ACCURATE SCORES according to the specified scales (0-6, 0-10, 0-8, 0-6)
2. OUTPUT EXCLUSIVELY IN JSON FORMAT - NO OTHER TEXT

MAXIMUM ATTENTION: The scores must be:

- NUMERICALLY CORRECT according to the defined scales
- JUSTIFIED WITH CONCRETE EVIDENCE from the paper
- MATHEMATICALLY CONSISTENT (correct total sum)

ABSOLUTE REQUIREMENT: The output MUST be ONLY the specified JSON - NO additional text before or after

EVALUATION CRITERIA [MANDATORY SCORES]

ATTENTION: Each score MUST adhere EXACTLY to the indicated numerical scales

1. RELEVANCE [0-6 POINTS - ADHERE TO THE SCALE]
   Evaluate:

- Consistency with the theme of psychological well-being
- Clear definition of the skill to be enhanced
- Appropriateness of the skill in relation to the identified problem
- Connection to empowerment, lifelong learning, and skill training

Scales:

- 6-5: Excellent relevance, innovative skill, strong theoretical connection
- 4: Good relevance, clear skill, evident connection
- 3-2: Sufficient relevance, skill not entirely clear
- 1-0: Poor relevance, skill undefined or inappropriate

1. **STRUCTURAL COHERENCE [0-10 POINTS - ADHERE TO THE SCALE]**
   Evaluate:

- Logical and sequential structure
- Coherence between objectives, methodology, and activities
- Clarity of exposition and completeness
- Gradual progression of the intervention plan

Scales:

- 10-9: Impeccable structure, crystal-clear logical flow
- 8-7: Very good coherence, clear structure
- 6-5: Good structure with some disconnections
- 4-3: Problematic structure, several inconsistencies
- 2-0: Confused or absent structure

1. **ORIGINALITY [0-8 POINTS - ADHERE TO THE SCALE]**
   Evaluate:

- Creativity in the proposed approach
- Innovation compared to standard interventions
- Original combination of methodologies
- Distinctive and personalized elements

Scales:

- 8-7: Highly original, authentic innovative ideas
- 6-5: Good originality, significant innovative elements
- 4-3: Moderate originality, some creative insights
- 2-1: Poor originality, conventional approaches
- 0: No originality, completely mundane

1. **FEASIBILITY [0-6 POINTS - ADHERE TO THE SCALE]**
   Evaluate:

- Practical feasibility of the intervention
- Concreteness of the operational steps
- Realistic consideration of resources, timing, and contexts
- Management of critical issues, sustainability

Scales:

- 6-5: Excellent feasibility, concrete and practicable proposal
- 4: Good feasibility with some practical uncertainties
- 3-2: Dubious feasibility, several questions
- 1-0: Not feasible, impracticable elements

QUALITY INDICATORS

Elements that increase the score:

- References to scientific literature
- Coherent integration of multiple methodologies
- Attention to the ethical dimension
- Consideration of the life cycle
- Structured outcomes evaluation plan
- Setting described and justified
- Awareness of psychological variables

Elements that penalize:

- Generality or superficiality
- Lack of theoretical foundations
- Vague or non-measurable objectives
- Activities not linked to objectives
- Unrealistic timelines
- Lack of ethical considerations

EVALUATION PROCESS

1. Preliminary Reading: Identify skill, target, and present sections
2. Criterion-Based Analysis: Compare with scales, identify evidence
3. Coherence Check: Verify consistency between scores and overall quality

JSON OUTPUT [MANDATORY FORMAT - ONLY ACCEPTED RESPONSE]

CRITICAL: Respond EXCLUSIVELY with the following JSON. NO text before or after.

VERIFY SCORES:

- Relevance: 0-6 points
- Coherence: 0-10 points
- Originality: 0-8 points
- Feasibility: 0-6 points
- MAXIMUM TOTAL: 30 points

MANDATORY JSON FORMAT:

```
{
 "evaluation": {
  "scores": {
   "pertinence": {
    "value": "[0–6]",
    "justification": "Specific 2–3 line rationale citing concrete elements from the paper"
   },
   "coherence": {
    "value": "[0–10]",
    "justification": "Specific 2–3 line rationale citing concrete elements from the paper"
   },
   "originality": {
```

```
      "value": "[0–8]",
      "justification": "Specific 2–3 line rationale citing concrete elements from the paper"
    },
    "feasibility": {
      "value": "[0–6]",
      "justification": "Specific 2–3 line rationale citing concrete elements from the paper"
    }
   },
   "total": {
    "value": "[sum of scores]/30",
    "quality_band": "[High (26–30) | Medium (20–25) | Low (<20)]"
   },
   "overall_comment": {
    "strengths": ["List of 2–3 main strengths"],
    "areas_for_improvement": ["List of 2–3 areas for improvement"],
    "summary": "General 3–4 line comment summarizing the overall evaluation"
   },
   "identified_skill": "Name of the main skill presented in the paper",
   "intervention_target": "Target audience or context of the proposed intervention"
 }
}
```

GUIDING PRINCIPLES

- Objectivity: Evaluate the content, not the writing style
- Constructiveness: Maintain a tone that encourages learning
- Balance: Be neither too harsh nor too lenient
- Appreciation: Acknowledge attempts at innovation, even if imperfect
- Contextualization: Consider that they are students, not expert professionals

FINAL INSTRUCTIONS [MANDATORY SCORE VERIFICATION]

ABSOLUTE PRIORITY: SCORES + JSON

1. VERIFY SCORING SCALES: Relevance(0-6), Coherence(0-10), Originality(0-8), Feasibility(0-6)
2. CALCULATE CORRECT TOTAL: Exact sum of the 4 scores (/30)
3. PROVIDE ONLY THE JSON - Zero additional text
4. JUSTIFY EACH SCORE with specific elements from the paper
5. CHECK THE MATH: Verify that the sum is correct

FINAL REMINDER: ONLY JSON + CORRECT SCORES ACCORDING TO THE SPECIFIED SCALES

ACCEPTED RESPONSE: EXCLUSIVELY THE JSON SPECIFIED ABOVE